\documentclass[twocolumn,  graphicx, amsmath,amssymb, superscriptaddress]{revtex4-1}
 \usepackage{graphicx}
 \usepackage{dcolumn}
 \usepackage{bm}
 \usepackage{natbib}

\begin{document}
  \title{Quantum Rod Emission Coupled to Plasmonic Lattice Resonances: A Collective Directional Source of Polarized Light }

  \author{S. R. K. Rodriguez}\email{s.rodriguez@amolf.nl}
  \address{Center for Nanophotonics, FOM Institute AMOLF, c/o Philips Research Laboratories, High Tech Campus 4, 5656 AE Eindhoven, The Netherlands.}

  \author{G. Lozano}
  \address{Center for Nanophotonics, FOM Institute AMOLF, c/o Philips Research Laboratories, High Tech Campus 4, 5656 AE Eindhoven, The Netherlands.}

  \author{M. A. Verschuuren}
  \address{ Philips Research Laboratories, High Tech Campus 4, 5656 AE Eindhoven, The Netherlands.}

  \author{R. Gomes}
  \address{Physics and Chemistry of Nanostructures, Center for Nano and Biophotonics, Ghent University.}

  \author{K. Lambert}
   \address{Physics and Chemistry of Nanostructures, Center for Nano and Biophotonics, Ghent University.}

  \author{B. De Geyter}
   \address{Physics and Chemistry of Nanostructures, Center for Nano and Biophotonics, Ghent University.}
  \address{Photonics Research Group, Department of Information Technology, Ghent University-IMEC
Sint-Pietersnieuwstraat 41, B-9000 Gent}

   \author{A. Hassinen}
   \address{Physics and Chemistry of Nanostructures, Center for Nano and Biophotonics, Ghent University.}

  \author{D. Van Thourhout}
  \address{Photonics Research Group, Department of Information Technology, Ghent University-IMEC
Sint-Pietersnieuwstraat 41, B-9000 Gent}

  \author{Z. Hens}
  \address{Physics and Chemistry of Nanostructures, Center for Nano and Biophotonics, Ghent University.}

 \author{J. G\'{o}mez Rivas}
  \address{Center for Nanophotonics, FOM Institute AMOLF, c/o Philips Research Laboratories, High Tech Campus 4, 5656 AE Eindhoven, The Netherlands.}
  \address{{COBRA Research Institute, Eindhoven University of Technology, P.O. Box 513, 5600 MB Eindhoven, The Netherlands}}

\date{\today}

\begin{abstract}
We demonstrate that an array of optical antennas may render a thin
layer of randomly oriented semiconductor nanocrystals into an
enhanced and highly directional source of polarized light. The array
sustains collective plasmonic lattice resonances which are in
spectral overlap with the emission of the nanocrystals over narrow
angular regions. Consequently, different photon energies of visible
light are enhanced and beamed into definite directions.
\end{abstract}

\maketitle

  The development of efficient and tunable (in photon energy, directionality, and polarization)
nanoscale light emitters is a central goal for nanophotonics.
Coupled semiconductor nanocrystal quantum emitters and metallic
nanostructures offer an ideal platform for this
purpose~\cite{Farahani&Hecht05, Curto2010, Polman2006}: the emission
energy can be tuned by varying the nanocrystal size due to quantum
confinement of charge carriers, while the emitted light can be
enhanced and controlled by structuring the metal to sustain surface
plasmon polaritons which are resonant with the emission. It has been
shown that Localized Surface Plasmon Resonances (LSPRs) in metallic
nanoparticles may lead to a strong confinement of optical radiation
into subwavelength volumes, resulting in a drastic modification of
the emission spectra~\cite{Feldman08}, and radiative decay
rates~\cite{Muskens07}, of emitters in this volume. However, such
strong effects depend on an accurate positioning of the emitter in
the region where the large electromagnetic enhancements occur. It is
possible to overcome this position dependance by means of collective
resonances in periodic arrays of metallic nanostructures. When a
diffraction order is radiating in the plane of the array, i.e., at a
Rayleigh anomaly condition, diffractive coupling of localized
surface plasmons leads to collective, lattice-induced resonances
known as Surface Lattice Resonances (SLRs)~\cite{Zou&Schatz04,
Crozier, Auguie&Barnes08, Kravets08, Vecchi09}. In contrast with
LSPRs which typically manifest as broad spectral features in
extinction with a flat angular dispersion, sharp and dispersive
features in extinction may result from the excitation of
SLRs~\cite{Vecchi09b}. The dispersive character of SLRs enables to
design the interaction of the associated  surface polaritons with
light emitters in the vicinity of the array such that a resonant
enhancement of the emission may take place over narrow spectral and
angular regions. Moreover, the SLR polaritons extend over tens of
microns~\cite{Vecchi09b}, making it possible to obtain a collective
enhancement of the emission from large volumes. In this Letter, we
demonstrate that SLRs in an array of silver nanoantennas drastically
modify the emission of a thin layer of randomly oriented CdSe/CdS
core/shell Quantum Rods (QRs), rendering an enhanced and highly
directional source of polarized light.

\begin{figure}
\centerline{\includegraphics[width=8.5cm]{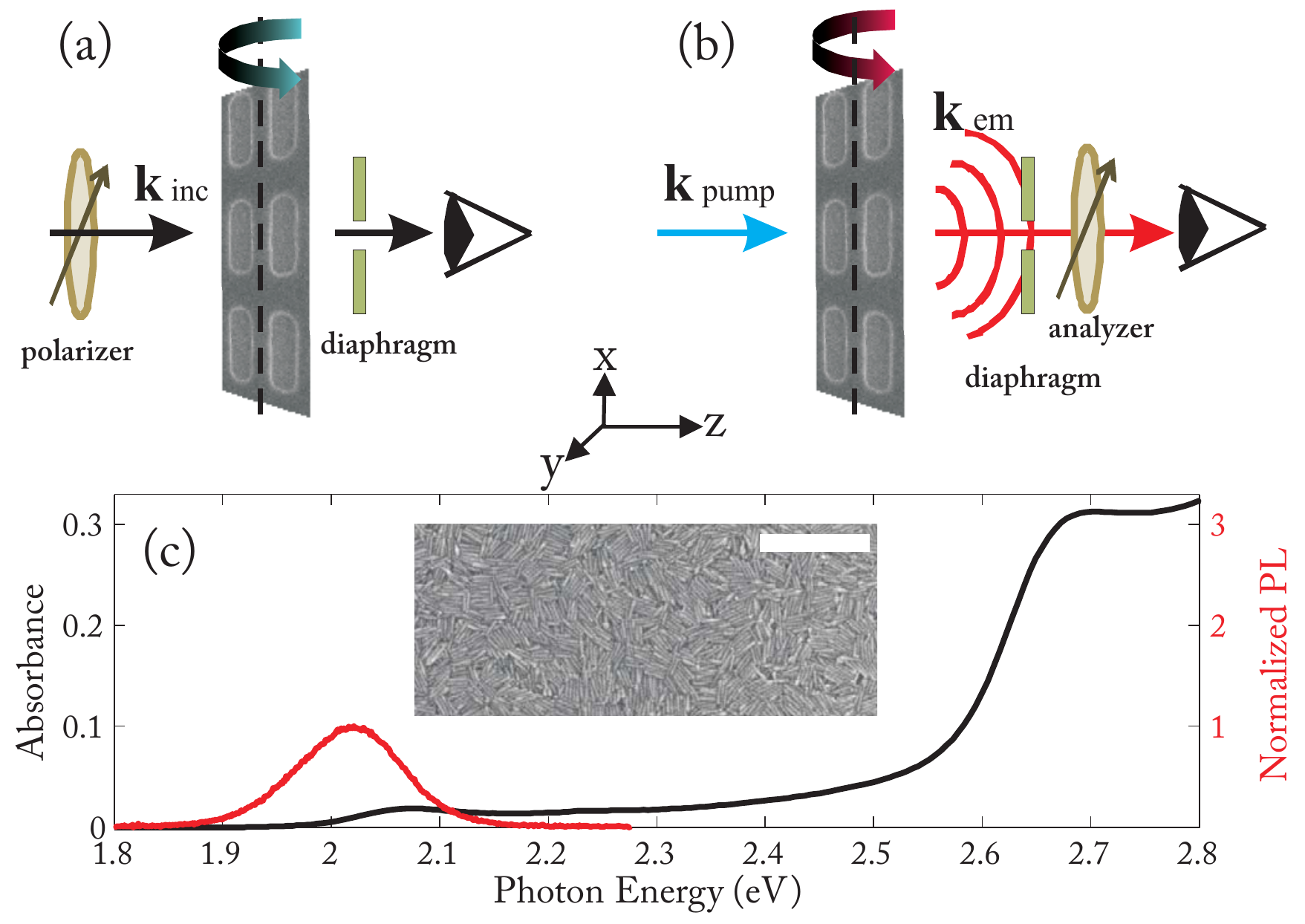}} \caption{Setup
used for (a) Extinction, and (b) PL enhancement measurements. The
sample is represented by a SEM image of a silver nanoantenna array,
with the dashed lines indicating the axis of rotation for the sample
in (a) and detector in (b). (c) Absorbance (left axis) and
normalized PL (right axis) of CdSe/CdS core/shell quantum rods
(QRs). The inset in (c) shows a SEM image of a spin-coated layer of
QRs; the scale bar denotes 200 nm.}\label{fig1}
\end{figure}

Silver nanoantenna arrays with a total size of $3 \times 3$ mm$^2$
were fabricated by Substrate Conformal Imprint Lithography
(SCIL)~\cite{SCIL} onto a fused silica substrate. A Scanning
Electron Microscope (SEM) image of an array is shown in Figs. 1(a)
and 1(b). This array has  antennas with dimensions $340 \times 110
\times 20$ nm$^3$ arranged in a lattice with constants $a_x = 500$
nm and $a_y = 200$ nm. A 20 nm layer of Si$_3$N$_4$ was deposited on
top of the array for a two-fold purpose: i) to passivate the silver,
and ii) to serve as a spacer layer between the antennas and the QRs
which prevents emission quenching~\cite{Wokaun83, Anger&Novotny06}.
The QRs were synthesized following Carbone $et$
$al$.~\cite{Carbone}. They have diameters and lengths of $4.0 \pm
0.6$ and $36.1 \pm 2.7$ nm, respectively,  and a Photoluminescence
(PL) quantum yield of $65\%$. The absorbance and PL spectra of the
QRs are shown in Fig. 1(c). We spin-coated a QR colloidal suspension
of ca 11 microM on top of the Si$_3$N$_4$ layer, forming a compact
layer of QRs [see inset of Fig. 1(c)] with a thickness of 60 nm.

\begin{figure}
\centerline{\includegraphics[width=8.5cm]{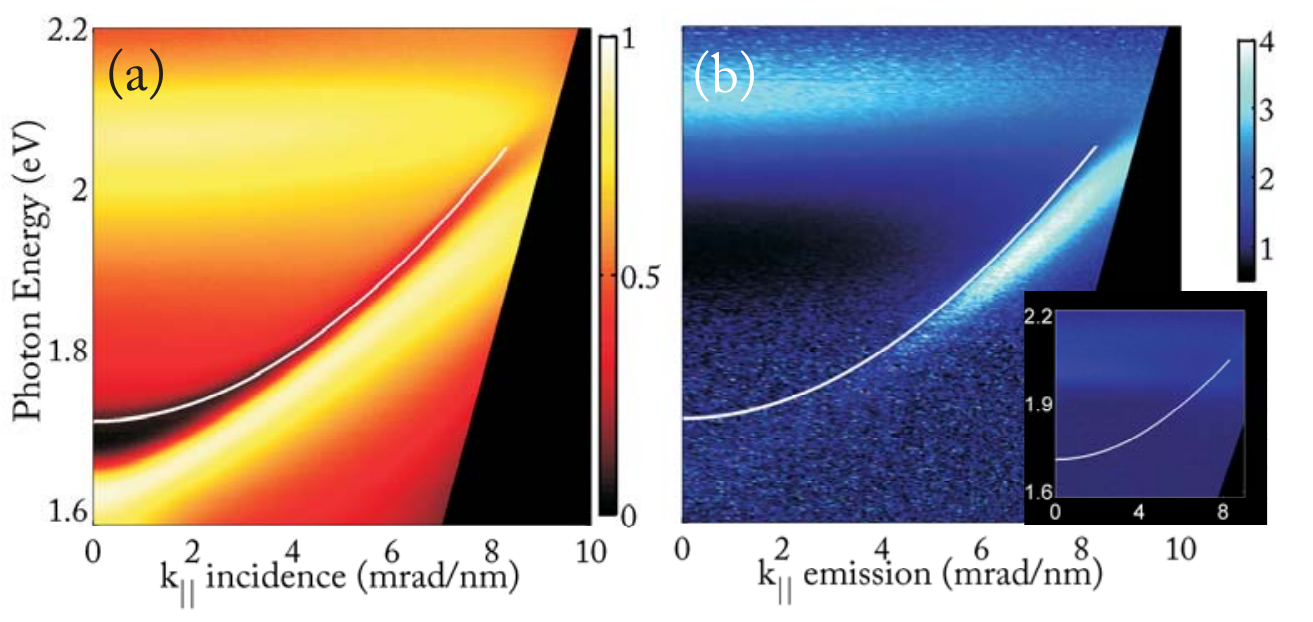}} \caption{(a)
Extinction and (b - main panel) PL Enhancement angular spectra of
p-polarized light of the sample described in the text. PL
Enhancement of s-polarized light is shown in the inset of (b). The
white lines indicate the degenerate ($\pm$1,0) Rayleigh
anomalies.}\label{fig2}
\end{figure}

We measured the variable angle zeroth-order extinction and
PhotoLuminescence Enhancement (PLE) spectra of the structure. For
the extinction measurements [see Fig. 1(a)], a p-polarized
collimated beam from a halogen lamp impinged onto the sample, which
we rotated around the x-axis while keeping the detector fixed. The
transmission through the array, $T_{in}$, was normalized to the
transmission outside the array (but through the substrate,
Si$_3$N$_4$, and QRs layers), $T_{out}$, to obtain the zeroth order
transmittance $T_0 = T_{in} / T_{out}$. The extinction, defined as
$1-T_0$, is shown in Fig. 2(a). The measurements in Fig. 2(a) are
shown as a function of the photon energy and the in-plane component
of the incident wave vector, $k_\| = k_0$ sin($\theta$)$\hat{y}$,
where $\theta$ is the angle of incidence and $k_0= \frac{2 \pi
}{\lambda_0}$ is the magnitude of the free space wave vector with
$\lambda_0$ the vacuum wavelength. For the PLE measurements, [see
Fig. 1(b)] the sample was excited by a continuous wave laser with an
energy of $2.81$ eV and an irradiance of $2$ mW/mm$^2$ at a fixed
angle of incidence $\theta = 5^{\circ}$. The pump irradiance was
confirmed to be far below saturation by measurements not shown here.
From the variable angle emission of the QRs inside the array
$I_{in}$, and outside the array $I_{out}$, we obtained a PLE  factor
given by $I_{in}/I_{out}$. Figure 2(b) shows the PLE dispersion
diagram for p-polarized emission in the main panel, and s-polarized
emission in the inset. The measurements in Fig. 2(b) are shown as a
function of the photon energy and the in-plane component of the wave
vector of the emitted light. The white lines in Figs. 2(a) and 2(b)
indicate the degenerate ($\pm$1,0) Rayleigh anomalies. They are
calculated from the conservation of the parallel component of the
wave vector, i.e., $k_{out}^2 = ( k_{x} \pm  m_1 G_x)^2 + ( k_{y}
\pm  m_2 G_y)^2$, with  $k_{out}$ the magnitude of the scattered
wave vector, $k_\| = (k_{x}, k_{y})$  the wave vector components
parallel to the surface, the integers ($m_1$, $m_2$) defining the
order of diffraction, and $\vec{G} = (G_x= \frac{2 \pi}{a_x}$,$G_y =
\frac{2 \pi}{a_y}$) the reciprocal lattice vector of the array. The
array of antennas was assumed to be embedded in a homogeneous medium
with n=1.45.

The broad peak in extinction near $2.07$ eV in Fig. 2(a) corresponds
to  the LSPR for the short axis of the nanoantennas. Notice that its
peak energy remains constant over the angular spectrum, which is a
manifestation of the non-dispersive character of localized
resonances.  The extinction also shows a pronounced dip near $1.69$
eV at $k_\| = 0$, shifting towards higher energies for an inclined
incidence. The good agreement between this dip
 and the calculated Rayleigh anomalies indicates that frequency dispersion of
the surrounding medium is negligible, since a constant refractive
index was assumed for the calculation. The coupling of the LSPR  to
the Rayleigh anomalies leads to narrow peaks in extinction on the
low energy side of the Rayleigh anomalies, corresponding to the
excitation of the ($\pm 1,0$) SLRs. As discussed in
Ref.~\cite{SRKR11a},  the peak energy, dispersion, and linewidth of
SLRs are determined by the coupling strength between the LSPR and
the Rayleigh anomalies. The salient feature of the present system is
that only until large values of $k_{||}$, the p-polarized SLRs cross
in energy with the emission bandwidth of the QRs. For this reason,
there is a resonant enhancement of the p-polarized emission by the
SLRs at large values of  $k_{||}$  only, as observed in the main
panel in Fig. 2(b). Moreover, the PLE attains a dispersive character
resembling the dispersion of the SLRs in extinction, i.e., the
enhancement factor in the region where the quantum dots emit is
roughly proportional to the extinction. The SLR enhanced emission is
absent for s polarization [inset of Fig. 2(b)], where only a very
weak feature (PLE $\sim 1.3$) near the Rayleigh anomaly condition is
observed. The contrasting enhancement for s and p polarization
arises from the plasmonic response of the nanoantennas, which
determines the polarization dependent excitation of SLRs. We stress
that although the Rayleigh anomalies have a purely geometrical
origin and are therefore polarization independent, the excitation of
the SLRs is determined by the shape and lattice dependent
polarizability of the nanoantennas~\cite{Zou&Schatz04}. The
different dimensions of the nanoantennas and the lattice along the x
and y directions give rise to SLRs in the extinction spectra which
overlap the emission of the QRs for p [see Fig. 2(a)], but not for s
(not shown here) polarization. Therefore, the structure acts as an
enhanced and directional source of p-polarized light.

In general, the total PLE may be factored into its contributions
from the pump and emission energies~\cite{Feldman08, Vecchi09}. We
have verified that the influence of resonant pump enhancements is
negligible by exciting the sample with different k-vectors and/or
polarizations. Moreover, the total enhancement factor may slightly
change for different pump conditions, but the features in the PLE
dispersion diagram remain unchanged because the emission is molded
by the dispersion of the SLRs at the emission energies. When
considering the maximum 4-fold enhancement herein reported, it
should be noted that we are starting with a fairly  high quantum
efficiency emitter. This condition makes large enhancement factors
less feasible~\cite{Weitz83}, but demonstrates the possibility to
improve and tailor the emission of an efficient light source.

\begin{figure}
\centerline{\includegraphics[width=9cm]{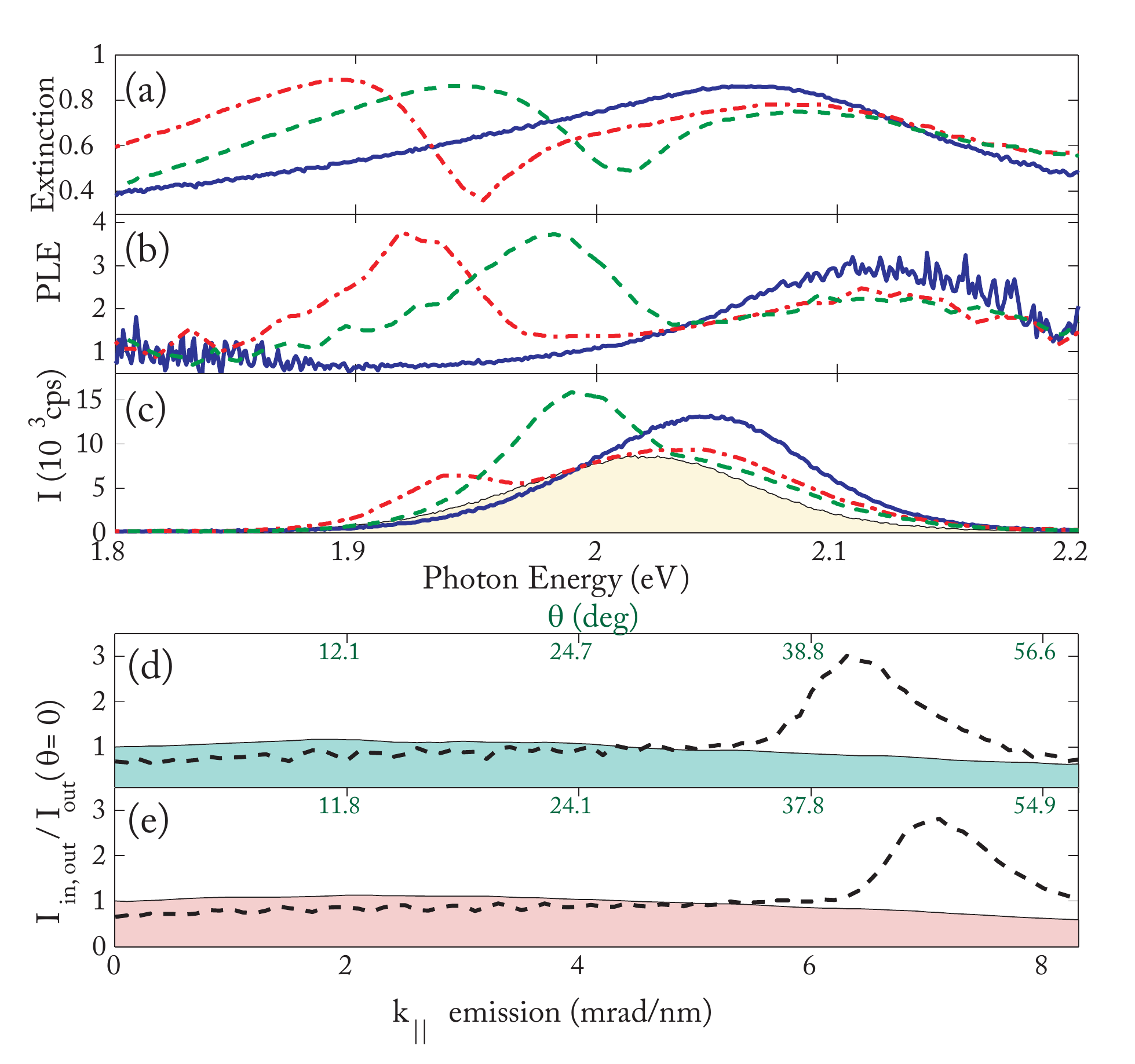}} \caption{(a)
Extinction, (b) PL Enhancement, and (c) emission inside the array,
i.e. $I_{in}$, in $10^3$ counts per second, at  $k_{||}= 0$ (blue
solid lines), $k_{||}=7$ mrad/nm (red dash-dot lines), and
$k_{||}=7.9$ mrad/nm (green dashed lines). The filled area in (c) is
the emission outside the array, i.e. $I_{out}$, at $k_{||}= 0$;
$I_{out}$ for the larger values of $k_{||}$  (not shown here) has
the same line shape as in $k_{||}= 0$, but with a lower amplitude.
(d) and (e) show  the angular dependent emission for an energy of
(d) $1.89$ eV, and (e) $1.93$ eV, with $I_{in}$  as black dashed
lines and $I_{out}$  as a filled area.   Both  $I_{in}$  and
$I_{out}$ are normalized to the value of $I_{out}$ at $\theta=0$ for
each energy.  The upper ticks indicate the value of the emission
angle $\theta$ corresponding to the $k_{||}$ values in the x-axis.
All measurements are shown for p-polarized light.}\label{fig3}
\end{figure}

 In Figs. 3(a) and 3(b) we show cuts of the spectra in Figs. 2(a)
and 2(b), with the blue solid, red dash-dotted, and green dashed
lines corresponding to $k_{||}= 0$, $k_{||}=7$ mrad/nm, and
$k_{||}=7.9$ mrad/nm, respectively. We observe that the emission is
enhanced in the spectral regions close to where the extinction is
highest, but the features appear to be shifted. We attribute this
shift to the different excitation conditions in the extinction and
PLE measurements. Namely, whereas the extinction corresponds to the
removal of energy from an incident plane wave, the PLE corresponds
to the out-coupled emission by the array when it is locally excited.
Differences in the Far Field (FF) and  Near Field (NF)  spectra have
been discussed in the context of emission ~\cite{Greffet2000} and
extinction of electromagnetic radiation~\cite{Bryant07, Lee,
Giannini10}. The extinction, which is determined by the FF
interference, can be spectrally shifted with respect to the NF due
to retardation effects. On the other hand, the loss of evanescent
modes present in the NF interaction between the emitters and the
antennas can change the FF emission spectra with respect to the NF
spectra.

To further elucidate the modification of the emission spectra, we
show in Fig. 3(c) the emission from the array, i.e. $I_{in}$, at the
same values of $k_{||} $ of Figs. 3(a) and 3(b); the filled area is
$I_{out}$ at $k_{||}=0$. For $k_{||}=0$, where only the LSPR
overlaps with the emission of the QRs, there is a blue shift of the
peak emission energy. In contrast, a red shift occurs for large
$k_{||}$, where the SLRs overlap with the emission of the QRs. In
view of the size polydispersion of the QRs which determines the
emission linewidth of the ensemble due to inhomogeneous broadening,
we attribute the blue shifted emission to a stronger interaction of
the smaller QRs with the LSPR, and the red shifted emission to a
stronger interaction of the bigger QRs with the SLRs. To asses the
angular modification of the emission by the array, we show in Figs.
3(d) and 3(e) $I_{in}$ (dashed lines) and $I_{out}$ (filled area) as
a function of $k_{||}$ for two photon energies: (d) $1.89$ eV and
(e) $1.93$ eV. Both $I_{in}$ and $I_{out}$ are normalized to the
forward emission outside the array, i.e., the value of $I_{out}$ at
$\theta=0$, for each energy. Outside the array the emission
resembles  a Lambertian emitter for both energies. Inside the array
the emission
 is suppressed at low values of $k_{||}$, indicating that resonant pump enhancement is negligible, since the emission is quenched in the absence of a resonant emission enhancement. At large values of $k_{||}$,
 the emission is enhanced within a narrow angular bandwidth for a given energy,
demonstrating the beaming of different emission energies into
different directions.

In conclusion, we have shown how collective resonances in
nanoantenna arrays enhance and modify  light emission from
semiconductor nanocrystals. By tailoring the spectral/angular
overlap between surface lattice resonances and the emission spectra,
we have demonstrated that a Lambertian emitter can be converted into
a directional source of polarized light.

This work was supported by the Netherlands Foundation Fundamenteel
Onderzoek der Materie (FOM) and the Nederlandse Organisatie voor
Wetenschappelijk Onderzoek (NWO), and is part of an industrial
partnership program between Philips and FOM. The work of S. R. K. R.
was partially funded by an Erasmus Mundus Masters in Photonics
scholarship.

\newpage


%

\end{document}